\documentclass[%
 reprint,
 amsmath,amssymb,
 aps,
prl,
]{revtex4-2}

\usepackage{xcolor}
\usepackage{graphicx}
\usepackage{dcolumn}
\usepackage{bm}

\begin{document}


\title{Recoil heating of a dielectric particle illuminated by a linearly polarized plane wave within the Rayleigh regime}

\author{Mohammad Ali Abbassi}
\affiliation{Department of Electrical Engineering, Sharif University of Technology, Tehran, Iran}
\email[]{abbassi\_ma@yahoo.com}
 




\date{\today}

\begin{abstract}
We investigate the recoil heating phenomenon experienced by a dielectric spherical particle when it interacts with a linearly polarized plane wave within the Rayleigh regime. We derive the fluctuating force acted upon the particle arising from the fluctuations of the electromagnetic fields. Our derivations reveal that the spectral density of the fluctuating force along the propagation direction is $7\hbar \omega_0 P_{\mathrm{scat}}/5c^2$. Meanwhile, along the direction of the electric and magnetic fields, it is $\hbar \omega_0 P_{\mathrm{scat}}/5c^2$ and $2\hbar \omega_0 P_{\mathrm{scat}}/5c^2$, respectively. Here, $P_{\mathrm{scat}}$ denotes the power scattered by the particle, $\hbar\omega_0$ represents the energy of a photon, and $c$ is the speed of light. Recoil heating imposes fundamental limitations in levitated optomechanics, constraining the minimum temperatures achievable in cooling processes, the coherence time of the system, and the sensitivity of force measurements.
\end{abstract}

\maketitle


The quantum fluctuations of electromagnetic fields induce a heating mechanism in particle motion, resulting from random momentum transfer during photon scattering\cite{jain2016direct,gonzalez2021levitodynamics,chang2010cavity,millen2020optomechanics}. This phenomenon, referred to as photon recoil heating, sets a fundamental limit on the achievable final temperature in cooling processes\cite{gieseler2012subkelvin,kamba2021recoil}. It also imposes constraints on the coherence time and force sensitivity in levitated optomechanical systems\cite{jain2016direct,tebbenjohanns2021quantum,gonzalez2023suppressing}. In addition, when a particle interacts with an electromagnetic field, it experiences damping of its motion, referred to as radiation damping\cite{abbassi2024radiation}.The equilibrium energy of the particle is determined by the balance between recoil heating and radiation damping, as dictated by the fluctuation-dissipation theorem\cite{jain2016direct,novotny2017radiation}.\par
Previous studies have estimated recoil heating by drawing an analogy with shot noise\cite{jain2016direct,seberson2020distribution}. In this letter, we present a rigorous derivation of recoil heating for a dielectric particle interacting with linearly polarized plane waves within the Rayleigh regime. We calculate the fluctuating force induced to the particle motion due to the quantum fluctuations of the electromagnetic fields. We utilize the quantization of electromagnetic fields within a dissipative medium, which is based on the principles of polarization and magnetization quantum noises\cite{milonni2019introduction,buhmann2013dispersion,buhmann2007dispersion}.
\par
Consider a dielectric spherical particle illuminated by a monochromatic plane wave. The plane wave is assumed to be x-polarized and propagating along the z-direction. The electromagnetic fields experienced by the particle can be written as:
\begin{subequations}
\begin{equation}\mathbf{\hat{E}}(\mathbf{r},t)=\frac{E_0}{2}\mathbf{e}_x e^{i(k_0 z-\omega_0 t)}+\int_0^\infty \mathbf{\hat{E}}_N(\mathbf{r},\omega)e^{-i\omega t}d\omega +\mathrm{H.c.},\end{equation}
\begin{equation}\begin{split}\mathbf{\hat{B}}(\mathbf{r},t)=&\frac{E_0}{2c} \mathbf{e}_y e^{i(k_0 z-\omega_0 t)}\\&+\int_0^\infty\frac{1}{i\omega} \mathbf{\nabla}\times \mathbf{\hat{E}}_N(\mathbf{r},\omega)e^{-i\omega t}d\omega +\mathrm{H.c.}.\end{split}\end{equation}
\end{subequations}
The first terms in the above expressions represent the incident plane wave, while the second terms are dedicated to describing the quantum fluctuations of the electromagnetic fields. Here, $\omega_0$ and $k_0$ represent the angular frequency and wave number of the incident wave, respectively, and $c$ denotes the speed of light.\par
We assume that the incident wave is in the classical limit, which implies that the fluctuations are identical to those for vacuum. Consequently, the electromagnetic noise is assumed to have a vanishing average and obeys the following correlation relations\cite{milonni2019introduction,buhmann2013dispersion,buhmann2007dispersion}:
\begin{subequations}
\begin{equation}\langle \mathbf{\hat{E}}_N(\mathbf{r},\omega)\mathbf{\hat{E}}_N^\dagger(\mathbf{r}^\prime,\omega^\prime) \rangle=\frac{\hbar\mu_0\omega^2}{\pi}\mathrm{Im}\left[\mathbf{G}_0(\mathbf{r},\mathbf{r}^\prime,\omega)\right]\delta(\omega-\omega^\prime),\end{equation}
\begin{equation}\begin{split}\langle \mathbf{\hat{E}}_N^\dagger(\mathbf{r},\omega)\mathbf{\hat{E}}_N(\mathbf{r}^\prime,\omega^\prime) \rangle&=\langle \mathbf{\hat{E}}_N(\mathbf{r},\omega)\mathbf{\hat{E}}_N(\mathbf{r}^\prime,\omega^\prime) \rangle\\&=\langle \mathbf{\hat{E}}_N^\dagger(\mathbf{r},\omega)\mathbf{\hat{E}}_N^\dagger(\mathbf{r}^\prime,\omega^\prime) \rangle=0.\end{split}\end{equation}
\end{subequations}
Here, $\mathbf{G}_0$ represents the dyadic Green's function of the free space\cite{novotny2012principles}.
\par
If the particle is much smaller than the wavelength of the incident wave, it can be modeled as an electric dipole\cite{chaumet2000time,abbassi2018inclusion,albaladejo2009scattering}. This equivalent dipole moment $\mathbf{\hat{p}}$ can be expressed by:
\begin{equation}\mathbf{\hat{p}}=\frac{E_0}{2} \alpha(\omega_0) \mathbf{e}_x e^{-i\omega_0 t}+\int_0^\infty  \alpha(\omega)\mathbf{\hat{E}}_N(\mathbf{r}_p,\omega)e^{-i\omega t}d\omega +\mathrm{H.c.},\end{equation}
where $\alpha(\omega)$ represents the polarizability of the particle, defined as:
\begin{equation}\alpha(\omega)=\frac{\alpha_0}{1-i\omega^3\alpha_0/6\pi\epsilon_0c^3}.\end{equation}
Here, $\alpha_0=4\pi\epsilon_0R_p^3(\epsilon_p-1)/(\epsilon_p+2)$ is the quasi-static polarizability of the particle,   with $R_p$ being the particle's radius and $\epsilon_p$ being its dielectric constant\cite{albaladejo2009scattering}.\par
The force exerted upon the particle can be calculated from\cite{novotny2012principles,chaumet2000time}:
\begin{equation}\label{eq:F}\mathbf{\hat{F}}=\left( \mathbf{\hat{p}}\cdot \mathbf{\nabla}\right) \mathbf{\hat{E}}+\frac{\partial \mathbf{\hat{p}}}{\partial t}\times \mathbf{\hat{B}}.\end{equation}
This force can be decomposed to two parts:
$\mathbf{\hat{F}}=\mathbf{\bar{F}}+\delta\mathbf{\hat{F}}$.
The first term, $\mathbf{\bar{F}}$, represents the deterministic part of the force exerted by the incident wave on the particle. Since the incident wave is a plane wave, $\mathbf{\bar{F}}$ is equivalent to the radiation pressure, which is given by:
\begin{equation}\mathbf{\bar{F}}=\frac{k_0E_0^2}{2}\mathrm{Im}\left[\alpha(\omega_0)\right]\mathbf{e}_z.\end{equation}
The second term, $\delta\mathbf{\hat{F}}$ stands for the fluctuating force acted upon the particle, which will be discussed in detail in the following.\par
{\it Fluctuating force along x-direction.---}
As indicated by Eq. \ref{eq:F}, the fluctuating force along the x-direction, which aligns with the polarization direction of the incident wave, is given by:
\begin{equation}\begin{split}\delta \hat{F}_x&=\bar{p}_x\int_0^\infty \partial_x\hat{E}_{N_x}(\mathrm{r}_p,\omega)e^{-i\omega t}d\omega\\&+ \partial_z\bar{E}_x \int_0^\infty \alpha(\omega) \hat{E}_{N_z}(\mathrm{r}_p,\omega)e^{-i\omega t}d\omega\\&+\bar{B}_y \int_0^\infty i\omega \alpha(\omega) \hat{E}_{N_z}(\mathrm{r}_p,\omega)e^{-i\omega t}d\omega \\&+ \mathrm{H.c.},\end{split}\end{equation}
where
\begin{subequations}
\begin{equation}\bar{p}_x=\mathrm{Re}\left[\alpha(\omega_0) E_0e^{-i\omega_0 t}\right],\end{equation}
\begin{equation}\partial_z\bar{E}_x=\mathrm{Re}\left[\frac{i\omega_0}{c}E_0e^{-i\omega_0 t}\right],\end{equation}
\begin{equation}\bar{B}_y=\mathrm{Re}\left[\frac{E_0}{c}e^{-i\omega_0 t}\right].\end{equation}
\end{subequations}
By utilizing the following relations regarding $\mathbf{G}_0$ \cite{SM}:
\begin{subequations}
\label{eq:G_relations}
\begin{equation}\mathrm{Im}\left[\mathbf{G}_0(\mathbf{r}_p,\mathbf{r}_p,\omega)\right]=\frac{\omega}{6\pi c}\mathbf{I},\end{equation}
\begin{equation}\mathrm{Im}\left[\partial_i\mathbf{G}_0(\mathbf{r}_p,\mathbf{r}_p,\omega)\right]=\mathrm{Im}\left[\partial_i^\prime\mathbf{G}_0(\mathbf{r}_p,\mathbf{r}_p,\omega)\right]=0,\end{equation}
\begin{equation}\mathrm{Im}\left[\partial_i\partial_j^\prime\mathbf{G}_0(\mathbf{r}_p,\mathbf{r}_p,\omega)\right]=\frac{\omega^3}{15\pi c^3}\delta_{ij}\mathbf{I}-\frac{\omega^3}{60\pi c^3}(\mathbf{e}_i\mathbf{e}_j+\mathbf{e}_j\mathbf{e}_i),\end{equation}
\end{subequations}
it can be easily shown that the auto-correlation function of $\delta \hat{F}_x$ can be expressed as\cite{SM}:
\begin{equation}\label{eq:cor_Fx}\begin{split}\langle \delta \hat{F}_x(t)&\delta \hat{F}_x(t+\tau) \rangle=\\ &\frac{\hbar\mu_0}{2\pi}\mathrm{Re}\left[|\alpha(\omega_0)|^2E_0^2 e^{i\omega_0\tau}\right]\int_0^\infty \frac{\omega^5}{30\pi c^3} e^{i\omega \tau}d\omega+\\&\frac{\hbar\mu_0}{2\pi}\mathrm{Re}\left[\frac{\omega_0^2}{c^2}E_0^2 e^{i\omega_0\tau}\right]\int_0^\infty \frac{\omega^3}{6\pi c}|\alpha(\omega)|^2 e^{i\omega \tau}d\omega-\\&\frac{\hbar\mu_0}{\pi}\mathrm{Re}\left[\frac{i\omega_0}{c^2}E_0^2 e^{i\omega_0\tau}\right]\int_0^\infty \frac{i\omega^4}{6\pi c}|\alpha(\omega)|^2 e^{i\omega \tau}d\omega +\\ &\frac{\hbar\mu_0}{2\pi}\mathrm{Re}\left[\frac{E_0^2}{c^2}e^{i\omega_0\tau}\right]\int_0^\infty \frac{\omega^5}{6\pi c}|\alpha(\omega)|^2 e^{i\omega \tau}d\omega.\end{split}\end{equation}
We can now determine the spectral density of $\delta \hat{F}_x$ by taking the Fourier transform of its auto-correlation function, defined as follows:
\begin{equation}S_{F_xF_x}(\Omega)=\int_{-\infty}^{+\infty} \langle \delta \hat{F}_x(t)\delta \hat{F}_x(t+\tau) \rangle e^{i\Omega \tau}d\tau,\end{equation}
resulting in
\begin{equation}\begin{split}S_{F_xF_x}(\Omega)=&\int_{0}^{+\infty}\int_{-\infty}^{+\infty}\frac{\hbar\mu_0}{4\pi}\alpha_0^2 E_0^2\times\\ &\bigg[\Big(\frac{\omega^5}{5\pi c^3}+\frac{\omega^3\omega_0^2}{6\pi c^3}+\frac{\omega^4\omega_0}{3\pi c^3}\Big)e^{i(\omega+\omega_0+\Omega)\tau}+\\&\Big(\frac{\omega^5}{5\pi c^3}+\frac{\omega^3\omega_0^2}{6\pi c^3}-\frac{\omega^4\omega_0}{3\pi c^3}\Big)e^{i(\omega-\omega_0+\Omega)\tau}\bigg]d\tau d\omega.\end{split}\end{equation}
It is important to note that we employ the approximation $|\alpha(\omega)|^2\simeq \alpha_0^2$ in deriving the above expression. Evaluating the integral over $\tau$ yields:
\begin{equation}\begin{split}S_{F_xF_x}(\Omega)=&\int_{0}^{+\infty}\frac{\hbar\mu_0}{2}\alpha_0^2 E_0^2\times\\&\bigg[\Big(\frac{\omega^5}{5\pi c^3}+\frac{\omega^3\omega_0^2}{6\pi c^3}+\frac{\omega^4\omega_0}{3\pi c^3}\Big)\delta(\omega+\omega_0+\Omega)+\\&\Big(\frac{\omega^5}{5\pi c^3}+\frac{\omega^3\omega_0^2}{6\pi c^3}-\frac{\omega^4\omega_0}{3\pi c^3}\Big)\delta(\omega-\omega_0+\Omega)\bigg] d\omega.\end{split}\end{equation}
Since the mechanical frequencies are much smaller than the optical frequencies, i.e. $\Omega\ll \omega_0$, the spectral density of $\delta \hat{F}_x$ can be simplified to:
\begin{equation}\label{eq:Sxx} S_{F_xF_x}(\Omega)\simeq\frac{\hbar\mu_0 \alpha_0^2 E_0^2\omega_0^5}{60\pi c^3}=\frac{1}{5}\frac{\hbar\omega_0}{c^2}P_{\mathrm{scat}},\end{equation}
where $P_{\mathrm{scat}}=\omega_0^4\alpha_0^2E_0^2/12\pi\epsilon_0c^3$ denotes the power scattered by the particle.
\par
{\it Fluctuating force along y-direction.---}
We aim to determine the fluctuating force along the y-direction, which corresponds to the direction of the incident magnetic field. According to Eq. \ref{eq:F}, the fluctuating force along the y-direction is given by:
\begin{equation}\begin{split}\delta \hat{F}_y&=\bar{p}_x\int_0^\infty \partial_x\hat{E}_{N_y}(\mathrm{r}_p,\omega)e^{-i\omega t}d\omega\\&- \frac{\partial \bar{p}_x}{\partial t} \int_0^\infty \frac{1}{i\omega} \left[\partial_x\hat{E}_{N_y}(\mathrm{r}_p,\omega)-\partial_y\hat{E}_{N_x}(\mathrm{r}_p,\omega)\right]e^{-i\omega t}d\omega\\&+ \mathrm{H.c.}.\end{split}\end{equation}
Upon utilizing the relations in Eq. \ref{eq:G_relations}, we can calculate the auto-correlation function of $\delta \hat{F}_y$, resulting in\cite{SM}
\begin{equation}\begin{split}\langle \delta \hat{F}_y(t)&\delta \hat{F}_y(t+\tau) \rangle=\\ &\frac{\hbar\mu_0}{2\pi}\mathrm{Re}\left[|\alpha(\omega_0)|^2E_0^2 e^{i\omega_0\tau}\right]\int_0^\infty \frac{\omega^5}{15\pi c^3} e^{i\omega \tau}d\omega-\\&\frac{\hbar\mu_0}{\pi}\mathrm{Re}\left[i\omega_0 |\alpha(\omega_0)|^2E_0^2 e^{i\omega_0\tau}\right]\int_0^\infty \frac{i\omega^4}{12\pi c^3} e^{i\omega \tau}d\omega +\\ &\frac{\hbar\mu_0}{2\pi}\mathrm{Re}\left[\omega_0^2|\alpha(\omega_0)|^2E_0^2e^{i\omega_0\tau}\right]\int_0^\infty \frac{\omega^3}{6\pi c^3} e^{i\omega \tau}d\omega.\end{split}\end{equation}
Then, we can derive the spectral density of $\delta \hat{F}_y$, as follows:
\begin{equation}\begin{split}S_{F_yF_y}(\Omega)=&\int_{0}^{+\infty}\int_{-\infty}^{+\infty}\frac{\hbar\mu_0}{4\pi}\alpha_0^2 E_0^2\times\\ &\bigg[\Big(\frac{\omega^5}{15\pi c^3}+\frac{\omega^4\omega_0}{6\pi c^3}+\frac{\omega^3\omega_0^2}{6\pi c^3}\Big)e^{i(\omega+\omega_0+\Omega)\tau}+\\&\Big(\frac{\omega^5}{15\pi c^3}-\frac{\omega^4\omega_0}{6\pi c^3}+\frac{\omega^3\omega_0^2}{6\pi c^3}\Big)e^{i(\omega-\omega_0+\Omega)\tau}\bigg]d\tau d\omega.\end{split}\end{equation}
If we perform the integration over $\tau$, one obtains:
\begin{equation}\begin{split}S_{F_yF_y}(\Omega)=&\int_{0}^{+\infty}\frac{\hbar\mu_0}{2}\alpha_0^2 E_0^2\times\\&\bigg[\Big(\frac{\omega^5}{15\pi c^3}+\frac{\omega^4\omega_0}{6\pi c^3}+\frac{\omega^3\omega_0^2}{6\pi c^3}\Big)\delta(\omega+\omega_0+\Omega)+\\&\Big(\frac{\omega^5}{15\pi c^3}-\frac{\omega^4\omega_0}{6\pi c^3}+\frac{\omega^3\omega_0^2}{6\pi c^3}\Big)\delta(\omega-\omega_0+\Omega)\bigg] d\omega.\end{split}\end{equation}
Eventually, we can simplify the spectral density of $\delta \hat{F}_y$ to
\begin{equation}S_{F_yF_y}(\Omega)\simeq\frac{\hbar\mu_0 \alpha_0^2 E_0^2\omega_0^5}{30\pi c^3}=\frac{2}{5}\frac{\hbar\omega_0}{c^2}P_{\mathrm{scat}},\end{equation}
given that the mechanical frequencies are much smaller than the optical frequency $\omega_0$.
\par
{\it Fluctuating force along z-direction.---}
We now aim to derive the fluctuating force along the propagation direction. As indicated by Eq. \ref{eq:F}, the fluctuating force along the z-direction is given by:
\begin{equation}\begin{split}\delta \hat{F}_z&=\bar{p}_x\int_0^\infty \partial_x\hat{E}_{N_z}(\mathrm{r}_p,\omega)e^{-i\omega t}d\omega\\&+ \frac{\partial \bar{p}_x}{\partial t} \int_0^\infty \frac{1}{i\omega} \left[\partial_z\hat{E}_{N_x}(\mathrm{r}_p,\omega)-\partial_x\hat{E}_{N_z}(\mathrm{r}_p,\omega)\right]e^{-i\omega t}d\omega\\&-\bar{B}_y \int_0^\infty i\omega \alpha(\omega) \hat{E}_{N_x}(\mathrm{r}_p,\omega)e^{-i\omega t}d\omega\\&+ \mathrm{H.c.}.\end{split}\end{equation}
Upon using the relations given in Eq. \ref{eq:G_relations}, the auto-correlation function of $\delta \hat{F}_z$ can be written as\cite{SM}:
\begin{equation}\begin{split}\langle \delta \hat{F}_z(t)&\delta \hat{F}_z(t+\tau) \rangle=\\ &\frac{\hbar\mu_0}{2\pi}\mathrm{Re}\left[|\alpha(\omega_0)|^2E_0^2 e^{i\omega_0\tau}\right]\int_0^\infty \frac{\omega^5}{15\pi c^3} e^{i\omega \tau}d\omega-\\&\frac{\hbar\mu_0}{\pi}\mathrm{Re}\left[i\omega_0 |\alpha(\omega_0)|^2E_0^2 e^{i\omega_0\tau}\right]\int_0^\infty \frac{i\omega^4}{12\pi c^3} e^{i\omega \tau}d\omega+\\ &\frac{\hbar\mu_0}{2\pi}\mathrm{Re}\left[\omega_0^2|\alpha(\omega_0)|^2E_0^2e^{i\omega_0\tau}\right]\int_0^\infty \frac{\omega^3}{6\pi c^3} e^{i\omega \tau}d\omega+\\&\frac{\hbar\mu_0}{2\pi}\mathrm{Re}\left[\frac{E_0^2}{c^2}e^{i\omega_0\tau}\right]\int_0^\infty |\alpha(\omega_0)|^2\frac{\omega^5}{6\pi c} e^{i\omega \tau}d\omega. \end{split}\end{equation}
Subsequently, we can derive the spectral density of $\delta \hat{F}_z$ by taking the Fourier transform from its auto-correlation function, resulting in
\begin{equation}\begin{split}S_{F_zF_z}(\Omega)=&\int_{0}^{+\infty}\int_{-\infty}^{+\infty}\frac{\hbar\mu_0}{4\pi}\alpha_0^2 E_0^2\times\\&\bigg[\Big(\frac{7\omega^5}{30\pi c^3}+\frac{\omega^4\omega_0}{6\pi c^3}+\frac{\omega^3\omega_0^2}{6\pi c^3}\Big)e^{i(\omega+\omega_0+\Omega)\tau}+\\&\Big(\frac{7\omega^5}{30\pi c^3}-\frac{\omega^4\omega_0}{6\pi c^3}+\frac{\omega^3\omega_0^2}{6\pi c^3}\Big)e^{i(\omega-\omega_0+\Omega)\tau}\bigg]d\tau d\omega.\end{split}\end{equation}
Upon evaluating the integral over $\tau$, one obtains
\begin{equation}\begin{split}S_{F_zF_z}(\Omega)=&\int_{0}^{+\infty}\frac{\hbar\mu_0}{2}\alpha_0^2 E_0^2\times\\&\bigg[\Big(\frac{7\omega^5}{30\pi c^3}+\frac{\omega^4\omega_0}{6\pi c^3}+\frac{\omega^3\omega_0^2}{6\pi c^3}\Big)\delta(\omega+\omega_0+\Omega)+\\&\Big(\frac{7\omega^5}{30\pi c^3}-\frac{\omega^4\omega_0}{6\pi c^3}+\frac{\omega^3\omega_0^2}{6\pi c^3}\Big)\delta(\omega-\omega_0+\Omega)\bigg] d\omega,\end{split}\end{equation}
which can be further simplified to
\begin{equation}S_{F_zF_z}(\Omega)\simeq\frac{7\hbar\mu_0 \alpha_0^2 E_0^2\omega_0^5}{60\pi c^3}=\frac{7}{5}\frac{\hbar\omega_0}{c^2}P_{\mathrm{scat}},\end{equation}
since the mechanical frequencies are much smaller than the optical frequency $\omega_0$.\par
{\it Particle dynamics.---} The interaction of the particle with the incident wave gives rise to a drag force known as radiation damping, along with a fluctuating force corresponding to recoil heating. Additionally, the presence of a thermal bath induces mechanical damping and thermal noise. Therefore, we can describe the particle motion along the x-direction using the equation:
\begin{equation}m\frac{d \hat{v}_x}{dt}=-m(\Gamma_m+\Gamma_x)\hat{v}_x+\hat{F}_x+\hat{\xi},\end{equation}
where $\hat{F}_x$ represents the fluctuating force, satisfying Eq. \ref{eq:Sxx}, and $\Gamma_x=P_{\mathrm{scat}}/mc^2$ is the radiation damping along x-direction\cite{abbassi2024radiation}. Furthermore,  $\Gamma_m$ represents the mechanical damping rate, and $\xi$ denotes the thermal noise, satisfying $\langle \xi(t)\xi(t')\rangle =2m\Gamma_mk_BT\delta(t-t')$. Here, $m$ denotes the particle mass, $k_B$ is the Boltzmann constant, and $T$ represents the ambient temperature\cite{milonni2019introduction,abbassi2019green}. We can now derive the variance of $v_x$, resulting in
\begin{equation}\langle v_x^2\rangle=\frac{2m\Gamma_mk_BT+\hbar\omega_0P_{\mathrm{scat}}/5c^2}{2m^2(\Gamma_m+P_{\mathrm{scat}}/mc^2)}.\end{equation}
When $\Gamma_m$ is much smaller than $P_{\mathrm{scat}}/mc^2$, which necessitates ultra-high vacuum, the variance of $v_x$ simplifies to $0.1\hbar\omega_0/m$. Under that condition, thermal decoherency is surpassed by recoil heating and radiation damping.  Similarly, one can obtain the variance of $v_y$ and $v_z$, that resulting in $0.2\hbar\omega_0/m$ and $7\hbar\omega_0/60m$, respectively. It should be noted that the radiation damping rate along y and z-directions are given by $\Gamma_y=P_{\mathrm{scat}}/mc^2$, and $\Gamma_z=6P_{\mathrm{scat}}/mc^2$, respectively\cite{abbassi2024radiation}.\par
The author is grateful to Lukas Novotny and Patrick Maurer for helpful discussions.

%



\begin{thebibliography}{20}%
\makeatletter
\providecommand \@ifxundefined [1]{%
 \@ifx{#1\undefined}
}%
\providecommand \@ifnum [1]{%
 \ifnum #1\expandafter \@firstoftwo
 \else \expandafter \@secondoftwo
 \fi
}%
\providecommand \@ifx [1]{%
 \ifx #1\expandafter \@firstoftwo
 \else \expandafter \@secondoftwo
 \fi
}%
\providecommand \natexlab [1]{#1}%
\providecommand \enquote  [1]{``#1''}%
\providecommand \bibnamefont  [1]{#1}%
\providecommand \bibfnamefont [1]{#1}%
\providecommand \citenamefont [1]{#1}%
\providecommand \href@noop [0]{\@secondoftwo}%
\providecommand \href [0]{\begingroup \@sanitize@url \@href}%
\providecommand \@href[1]{\@@startlink{#1}\@@href}%
\providecommand \@@href[1]{\endgroup#1\@@endlink}%
\providecommand \@sanitize@url [0]{\catcode `\\12\catcode `\$12\catcode
  `\&12\catcode `\#12\catcode `\^12\catcode `\_12\catcode `\%12\relax}%
\providecommand \@@startlink[1]{}%
\providecommand \@@endlink[0]{}%
\providecommand \url  [0]{\begingroup\@sanitize@url \@url }%
\providecommand \@url [1]{\endgroup\@href {#1}{\urlprefix }}%
\providecommand \urlprefix  [0]{URL }%
\providecommand \Eprint [0]{\href }%
\providecommand \doibase [0]{https://doi.org/}%
\providecommand \selectlanguage [0]{\@gobble}%
\providecommand \bibinfo  [0]{\@secondoftwo}%
\providecommand \bibfield  [0]{\@secondoftwo}%
\providecommand \translation [1]{[#1]}%
\providecommand \BibitemOpen [0]{}%
\providecommand \bibitemStop [0]{}%
\providecommand \bibitemNoStop [0]{.\EOS\space}%
\providecommand \EOS [0]{\spacefactor3000\relax}%
\providecommand \BibitemShut  [1]{\csname bibitem#1\endcsname}%
\let\auto@bib@innerbib\@empty
\bibitem [{\citenamefont {Jain}\ \emph {et~al.}(2016)\citenamefont {Jain},
  \citenamefont {Gieseler}, \citenamefont {Moritz}, \citenamefont {Dellago},
  \citenamefont {Quidant},\ and\ \citenamefont {Novotny}}]{jain2016direct}%
  \BibitemOpen
  \bibfield  {author} {\bibinfo {author} {\bibfnamefont {V.}~\bibnamefont
  {Jain}}, \bibinfo {author} {\bibfnamefont {J.}~\bibnamefont {Gieseler}},
  \bibinfo {author} {\bibfnamefont {C.}~\bibnamefont {Moritz}}, \bibinfo
  {author} {\bibfnamefont {C.}~\bibnamefont {Dellago}}, \bibinfo {author}
  {\bibfnamefont {R.}~\bibnamefont {Quidant}},\ and\ \bibinfo {author}
  {\bibfnamefont {L.}~\bibnamefont {Novotny}},\ }\bibfield  {title} {\bibinfo
  {title} {Direct measurement of photon recoil from a levitated nanoparticle},\
  }\href@noop {} {\bibfield  {journal} {\bibinfo  {journal} {Physical review
  letters}\ }\textbf {\bibinfo {volume} {116}},\ \bibinfo {pages} {243601}
  (\bibinfo {year} {2016})}\BibitemShut {NoStop}%
\bibitem [{\citenamefont {Gonzalez-Ballestero}\ \emph
  {et~al.}(2021)\citenamefont {Gonzalez-Ballestero}, \citenamefont
  {Aspelmeyer}, \citenamefont {Novotny}, \citenamefont {Quidant},\ and\
  \citenamefont {Romero-Isart}}]{gonzalez2021levitodynamics}%
  \BibitemOpen
  \bibfield  {author} {\bibinfo {author} {\bibfnamefont {C.}~\bibnamefont
  {Gonzalez-Ballestero}}, \bibinfo {author} {\bibfnamefont {M.}~\bibnamefont
  {Aspelmeyer}}, \bibinfo {author} {\bibfnamefont {L.}~\bibnamefont {Novotny}},
  \bibinfo {author} {\bibfnamefont {R.}~\bibnamefont {Quidant}},\ and\ \bibinfo
  {author} {\bibfnamefont {O.}~\bibnamefont {Romero-Isart}},\ }\bibfield
  {title} {\bibinfo {title} {Levitodynamics: Levitation and control of
  microscopic objects in vacuum},\ }\href@noop {} {\bibfield  {journal}
  {\bibinfo  {journal} {Science}\ }\textbf {\bibinfo {volume} {374}},\ \bibinfo
  {pages} {eabg3027} (\bibinfo {year} {2021})}\BibitemShut {NoStop}%
\bibitem [{\citenamefont {Chang}\ \emph {et~al.}(2010)\citenamefont {Chang},
  \citenamefont {Regal}, \citenamefont {Papp}, \citenamefont {Wilson},
  \citenamefont {Ye}, \citenamefont {Painter}, \citenamefont {Kimble},\ and\
  \citenamefont {Zoller}}]{chang2010cavity}%
  \BibitemOpen
  \bibfield  {author} {\bibinfo {author} {\bibfnamefont {D.~E.}\ \bibnamefont
  {Chang}}, \bibinfo {author} {\bibfnamefont {C.}~\bibnamefont {Regal}},
  \bibinfo {author} {\bibfnamefont {S.}~\bibnamefont {Papp}}, \bibinfo {author}
  {\bibfnamefont {D.}~\bibnamefont {Wilson}}, \bibinfo {author} {\bibfnamefont
  {J.}~\bibnamefont {Ye}}, \bibinfo {author} {\bibfnamefont {O.}~\bibnamefont
  {Painter}}, \bibinfo {author} {\bibfnamefont {H.~J.}\ \bibnamefont
  {Kimble}},\ and\ \bibinfo {author} {\bibfnamefont {P.}~\bibnamefont
  {Zoller}},\ }\bibfield  {title} {\bibinfo {title} {Cavity opto-mechanics
  using an optically levitated nanosphere},\ }\href@noop {} {\bibfield
  {journal} {\bibinfo  {journal} {Proceedings of the National Academy of
  Sciences}\ }\textbf {\bibinfo {volume} {107}},\ \bibinfo {pages} {1005}
  (\bibinfo {year} {2010})}\BibitemShut {NoStop}%
\bibitem [{\citenamefont {Millen}\ \emph {et~al.}(2020)\citenamefont {Millen},
  \citenamefont {Monteiro}, \citenamefont {Pettit},\ and\ \citenamefont
  {Vamivakas}}]{millen2020optomechanics}%
  \BibitemOpen
  \bibfield  {author} {\bibinfo {author} {\bibfnamefont {J.}~\bibnamefont
  {Millen}}, \bibinfo {author} {\bibfnamefont {T.~S.}\ \bibnamefont
  {Monteiro}}, \bibinfo {author} {\bibfnamefont {R.}~\bibnamefont {Pettit}},\
  and\ \bibinfo {author} {\bibfnamefont {A.~N.}\ \bibnamefont {Vamivakas}},\
  }\bibfield  {title} {\bibinfo {title} {Optomechanics with levitated
  particles},\ }\href@noop {} {\bibfield  {journal} {\bibinfo  {journal}
  {Reports on Progress in Physics}\ }\textbf {\bibinfo {volume} {83}},\
  \bibinfo {pages} {026401} (\bibinfo {year} {2020})}\BibitemShut {NoStop}%
\bibitem [{\citenamefont {Gieseler}\ \emph {et~al.}(2012)\citenamefont
  {Gieseler}, \citenamefont {Deutsch}, \citenamefont {Quidant},\ and\
  \citenamefont {Novotny}}]{gieseler2012subkelvin}%
  \BibitemOpen
  \bibfield  {author} {\bibinfo {author} {\bibfnamefont {J.}~\bibnamefont
  {Gieseler}}, \bibinfo {author} {\bibfnamefont {B.}~\bibnamefont {Deutsch}},
  \bibinfo {author} {\bibfnamefont {R.}~\bibnamefont {Quidant}},\ and\ \bibinfo
  {author} {\bibfnamefont {L.}~\bibnamefont {Novotny}},\ }\bibfield  {title}
  {\bibinfo {title} {Subkelvin parametric feedback cooling of a laser-trapped
  nanoparticle},\ }\href@noop {} {\bibfield  {journal} {\bibinfo  {journal}
  {Physical review letters}\ }\textbf {\bibinfo {volume} {109}},\ \bibinfo
  {pages} {103603} (\bibinfo {year} {2012})}\BibitemShut {NoStop}%
\bibitem [{\citenamefont {Kamba}\ \emph {et~al.}(2021)\citenamefont {Kamba},
  \citenamefont {Kiuchi}, \citenamefont {Yotsuya},\ and\ \citenamefont
  {Aikawa}}]{kamba2021recoil}%
  \BibitemOpen
  \bibfield  {author} {\bibinfo {author} {\bibfnamefont {M.}~\bibnamefont
  {Kamba}}, \bibinfo {author} {\bibfnamefont {H.}~\bibnamefont {Kiuchi}},
  \bibinfo {author} {\bibfnamefont {T.}~\bibnamefont {Yotsuya}},\ and\ \bibinfo
  {author} {\bibfnamefont {K.}~\bibnamefont {Aikawa}},\ }\bibfield  {title}
  {\bibinfo {title} {Recoil-limited feedback cooling of single nanoparticles
  near the ground state in an optical lattice},\ }\href@noop {} {\bibfield
  {journal} {\bibinfo  {journal} {Physical Review A}\ }\textbf {\bibinfo
  {volume} {103}},\ \bibinfo {pages} {L051701} (\bibinfo {year}
  {2021})}\BibitemShut {NoStop}%
\bibitem [{\citenamefont {Tebbenjohanns}\ \emph {et~al.}(2021)\citenamefont
  {Tebbenjohanns}, \citenamefont {Mattana}, \citenamefont {Rossi},
  \citenamefont {Frimmer},\ and\ \citenamefont
  {Novotny}}]{tebbenjohanns2021quantum}%
  \BibitemOpen
  \bibfield  {author} {\bibinfo {author} {\bibfnamefont {F.}~\bibnamefont
  {Tebbenjohanns}}, \bibinfo {author} {\bibfnamefont {M.~L.}\ \bibnamefont
  {Mattana}}, \bibinfo {author} {\bibfnamefont {M.}~\bibnamefont {Rossi}},
  \bibinfo {author} {\bibfnamefont {M.}~\bibnamefont {Frimmer}},\ and\ \bibinfo
  {author} {\bibfnamefont {L.}~\bibnamefont {Novotny}},\ }\bibfield  {title}
  {\bibinfo {title} {Quantum control of a nanoparticle optically levitated in
  cryogenic free space},\ }\href@noop {} {\bibfield  {journal} {\bibinfo
  {journal} {Nature}\ }\textbf {\bibinfo {volume} {595}},\ \bibinfo {pages}
  {378} (\bibinfo {year} {2021})}\BibitemShut {NoStop}%
\bibitem [{\citenamefont {Gonzalez-Ballestero}\ \emph
  {et~al.}(2023)\citenamefont {Gonzalez-Ballestero}, \citenamefont
  {Zieli{\'n}ska}, \citenamefont {Rossi}, \citenamefont {Militaru},
  \citenamefont {Frimmer}, \citenamefont {Novotny}, \citenamefont {Maurer},\
  and\ \citenamefont {Romero-Isart}}]{gonzalez2023suppressing}%
  \BibitemOpen
  \bibfield  {author} {\bibinfo {author} {\bibfnamefont {C.}~\bibnamefont
  {Gonzalez-Ballestero}}, \bibinfo {author} {\bibfnamefont {J.~A.}\
  \bibnamefont {Zieli{\'n}ska}}, \bibinfo {author} {\bibfnamefont
  {M.}~\bibnamefont {Rossi}}, \bibinfo {author} {\bibfnamefont
  {A.}~\bibnamefont {Militaru}}, \bibinfo {author} {\bibfnamefont
  {M.}~\bibnamefont {Frimmer}}, \bibinfo {author} {\bibfnamefont
  {L.}~\bibnamefont {Novotny}}, \bibinfo {author} {\bibfnamefont
  {P.}~\bibnamefont {Maurer}},\ and\ \bibinfo {author} {\bibfnamefont
  {O.}~\bibnamefont {Romero-Isart}},\ }\bibfield  {title} {\bibinfo {title}
  {Suppressing recoil heating in levitated optomechanics using squeezed
  light},\ }\href@noop {} {\bibfield  {journal} {\bibinfo  {journal} {PRX
  Quantum}\ }\textbf {\bibinfo {volume} {4}},\ \bibinfo {pages} {030331}
  (\bibinfo {year} {2023})}\BibitemShut {NoStop}%
\bibitem [{\citenamefont {Abbassi}\ and\ \citenamefont
  {Novotny}(2024)}]{abbassi2024radiation}%
  \BibitemOpen
  \bibfield  {author} {\bibinfo {author} {\bibfnamefont {M.~A.}\ \bibnamefont
  {Abbassi}}\ and\ \bibinfo {author} {\bibfnamefont {L.}~\bibnamefont
  {Novotny}},\ }\bibfield  {title} {\bibinfo {title} {Radiation damping of a
  rayleigh scatterer illuminated by a linearly polarized plane wave},\
  }\href@noop {} {\bibfield  {journal} {\bibinfo  {journal} {Physical Review
  A}\ }\textbf {\bibinfo {volume} {109}},\ \bibinfo {pages} {043531} (\bibinfo
  {year} {2024})}\BibitemShut {NoStop}%
\bibitem [{\citenamefont {Novotny}(2017)}]{novotny2017radiation}%
  \BibitemOpen
  \bibfield  {author} {\bibinfo {author} {\bibfnamefont {L.}~\bibnamefont
  {Novotny}},\ }\bibfield  {title} {\bibinfo {title} {Radiation damping of a
  polarizable particle},\ }\href@noop {} {\bibfield  {journal} {\bibinfo
  {journal} {Physical Review A}\ }\textbf {\bibinfo {volume} {96}},\ \bibinfo
  {pages} {032108} (\bibinfo {year} {2017})}\BibitemShut {NoStop}%
\bibitem [{\citenamefont {Seberson}\ and\ \citenamefont
  {Robicheaux}(2020)}]{seberson2020distribution}%
  \BibitemOpen
  \bibfield  {author} {\bibinfo {author} {\bibfnamefont {T.}~\bibnamefont
  {Seberson}}\ and\ \bibinfo {author} {\bibfnamefont {F.}~\bibnamefont
  {Robicheaux}},\ }\bibfield  {title} {\bibinfo {title} {Distribution of laser
  shot-noise energy delivered to a levitated nanoparticle},\ }\href@noop {}
  {\bibfield  {journal} {\bibinfo  {journal} {Physical Review A}\ }\textbf
  {\bibinfo {volume} {102}},\ \bibinfo {pages} {033505} (\bibinfo {year}
  {2020})}\BibitemShut {NoStop}%
\bibitem [{\citenamefont {Milonni}(2019)}]{milonni2019introduction}%
  \BibitemOpen
  \bibfield  {author} {\bibinfo {author} {\bibfnamefont {P.~W.}\ \bibnamefont
  {Milonni}},\ }\href@noop {} {\emph {\bibinfo {title} {An introduction to
  quantum optics and quantum fluctuations}}}\ (\bibinfo  {publisher} {Oxford
  University Press},\ \bibinfo {year} {2019})\BibitemShut {NoStop}%
\bibitem [{\citenamefont {Buhmann}(2013)}]{buhmann2013dispersion}%
  \BibitemOpen
  \bibfield  {author} {\bibinfo {author} {\bibfnamefont {S.~Y.}\ \bibnamefont
  {Buhmann}},\ }\href@noop {} {\emph {\bibinfo {title} {Dispersion Forces I:
  Macroscopic quantum electrodynamics and ground-state Casimir, Casimir--Polder
  and van der Waals forces}}},\ Vol.\ \bibinfo {volume} {247}\ (\bibinfo
  {publisher} {Springer},\ \bibinfo {year} {2013})\BibitemShut {NoStop}%
\bibitem [{\citenamefont {Buhmann}\ and\ \citenamefont
  {Welsch}(2007)}]{buhmann2007dispersion}%
  \BibitemOpen
  \bibfield  {author} {\bibinfo {author} {\bibfnamefont {S.~Y.}\ \bibnamefont
  {Buhmann}}\ and\ \bibinfo {author} {\bibfnamefont {D.-G.}\ \bibnamefont
  {Welsch}},\ }\bibfield  {title} {\bibinfo {title} {Dispersion forces in
  macroscopic quantum electrodynamics},\ }\href@noop {} {\bibfield  {journal}
  {\bibinfo  {journal} {Progress in quantum electronics}\ }\textbf {\bibinfo
  {volume} {31}},\ \bibinfo {pages} {51} (\bibinfo {year} {2007})}\BibitemShut
  {NoStop}%
\bibitem [{\citenamefont {Novotny}\ and\ \citenamefont
  {Hecht}(2012)}]{novotny2012principles}%
  \BibitemOpen
  \bibfield  {author} {\bibinfo {author} {\bibfnamefont {L.}~\bibnamefont
  {Novotny}}\ and\ \bibinfo {author} {\bibfnamefont {B.}~\bibnamefont
  {Hecht}},\ }\href@noop {} {\emph {\bibinfo {title} {Principles of
  nano-optics}}}\ (\bibinfo  {publisher} {Cambridge university press},\
  \bibinfo {year} {2012})\BibitemShut {NoStop}%
\bibitem [{\citenamefont {Chaumet}\ and\ \citenamefont
  {Nieto-Vesperinas}(2000)}]{chaumet2000time}%
  \BibitemOpen
  \bibfield  {author} {\bibinfo {author} {\bibfnamefont {P.~C.}\ \bibnamefont
  {Chaumet}}\ and\ \bibinfo {author} {\bibfnamefont {M.}~\bibnamefont
  {Nieto-Vesperinas}},\ }\bibfield  {title} {\bibinfo {title} {Time-averaged
  total force on a dipolar sphere in an electromagnetic field},\ }\href@noop {}
  {\bibfield  {journal} {\bibinfo  {journal} {Optics letters}\ }\textbf
  {\bibinfo {volume} {25}},\ \bibinfo {pages} {1065} (\bibinfo {year}
  {2000})}\BibitemShut {NoStop}%
\bibitem [{\citenamefont {Abbassi}\ and\ \citenamefont
  {Mehrany}(2018)}]{abbassi2018inclusion}%
  \BibitemOpen
  \bibfield  {author} {\bibinfo {author} {\bibfnamefont {M.~A.}\ \bibnamefont
  {Abbassi}}\ and\ \bibinfo {author} {\bibfnamefont {K.}~\bibnamefont
  {Mehrany}},\ }\bibfield  {title} {\bibinfo {title} {Inclusion of the
  backaction term in the total optical force exerted upon rayleigh particles in
  nonresonant structures},\ }\href@noop {} {\bibfield  {journal} {\bibinfo
  {journal} {Physical Review A}\ }\textbf {\bibinfo {volume} {98}},\ \bibinfo
  {pages} {013806} (\bibinfo {year} {2018})}\BibitemShut {NoStop}%
\bibitem [{\citenamefont {Albaladejo}\ \emph {et~al.}(2009)\citenamefont
  {Albaladejo}, \citenamefont {Marqu{\'e}s}, \citenamefont {Laroche},\ and\
  \citenamefont {S{\'a}enz}}]{albaladejo2009scattering}%
  \BibitemOpen
  \bibfield  {author} {\bibinfo {author} {\bibfnamefont {S.}~\bibnamefont
  {Albaladejo}}, \bibinfo {author} {\bibfnamefont {M.~I.}\ \bibnamefont
  {Marqu{\'e}s}}, \bibinfo {author} {\bibfnamefont {M.}~\bibnamefont
  {Laroche}},\ and\ \bibinfo {author} {\bibfnamefont {J.~J.}\ \bibnamefont
  {S{\'a}enz}},\ }\bibfield  {title} {\bibinfo {title} {Scattering forces from
  the curl of the spin angular momentum of a light field},\ }\href@noop {}
  {\bibfield  {journal} {\bibinfo  {journal} {Physical review letters}\
  }\textbf {\bibinfo {volume} {102}},\ \bibinfo {pages} {113602} (\bibinfo
  {year} {2009})}\BibitemShut {NoStop}%
\bibitem [{SM()}]{SM}%
  \BibitemOpen
  \href@noop {} {}\bibinfo {note} {See Supplemental Material.}\BibitemShut
  {Stop}%
\bibitem [{\citenamefont {Abbassi}\ and\ \citenamefont
  {Mehrany}(2019)}]{abbassi2019green}%
  \BibitemOpen
  \bibfield  {author} {\bibinfo {author} {\bibfnamefont {M.~A.}\ \bibnamefont
  {Abbassi}}\ and\ \bibinfo {author} {\bibfnamefont {K.}~\bibnamefont
  {Mehrany}},\ }\bibfield  {title} {\bibinfo {title} {Green's-function
  formulation for studying the backaction cooling of a levitated nanosphere in
  an arbitrary structure},\ }\href@noop {} {\bibfield  {journal} {\bibinfo
  {journal} {Physical Review A}\ }\textbf {\bibinfo {volume} {100}},\ \bibinfo
  {pages} {023823} (\bibinfo {year} {2019})}\BibitemShut {NoStop}%
\end{thebibliography}

\end{document}


\renewcommand{\theequation}{S.\arabic{equation}}

\title{Supplemental Material: Recoil heating of a dielectric particle illuminated by a linearly polarized plane wave within the Rayleigh regime}

\author{Mohammad Ali Abbassi}
\affiliation{Department of Electrical Engineering, Sharif University of Technology, Tehran, Iran}
\email[]{abbassi\_ma@yahoo.com}
 




\date{\today}

\maketitle

Here, we provide more details on some of the derivations in the main text.
\subsection{Dyadic Green's function relations}
The dyadic Green's function in the free space is given by 
\begin{equation}\mathbf{G}_0(\mathbf{r},\mathbf{r}^\prime;\omega)=\frac{e^{ik R}}{4\pi R}\Bigg[\left(1+\frac{i}{k R}-\frac{1}{k^2R^2}\right)\mathbf{I}+\left(\frac{3}{k^2R^2}-\frac{3i}{k R}-1\right)\frac{\mathbf{R}\mathbf{R}}{R^2}\Bigg],\end{equation}
where $k=\omega/c$ is the wave number, $\mathbf{I}$ denotes unit dyadic, and $\mathbf{R}=\mathbf{r}-\mathbf{r}^\prime$ represents the displacement vector from the source point $\mathbf{r}^\prime$ to the field point $\mathbf{r}$. By expanding $exp[ikR]$ into a series, it can be easily shown that the imaginary part of $\mathbf{G}_0$ is
\begin{equation}\label{eq:Gi}\mathrm{Im}\left[\mathbf{G}_0(\mathbf{r},\mathbf{r}^\prime;\omega)\right]=\frac{k}{4\pi}\left[\frac{2}{3}-\frac{2}{15}k^2R^2+\cdots\right]\mathbf{I}+\frac{k^3}{4\pi}\left[\frac{1}{15}-\frac{1}{210}k^2R^2+\cdots\right]\mathbf{R}\mathbf{R}.\end{equation}
We can now obtain the imaginary part of $\mathbf{G}_0$ at the source point by setting $R=0$, resulting in:
\begin{equation}\mathrm{Im}\left[\mathbf{G}_0(\mathbf{r}_p,\mathbf{r}_p;\omega)\right]=\frac{k}{6\pi}\mathbf{I}.\end{equation}
To obtain the derivatives of $\mathbf{G}_0$, we can apply $\partial_i$ on both sides of Eq. \ref{eq:Gi} that yields
\begin{equation}\label{eq:dGi}\mathrm{Im}\left[\partial_i\mathbf{G}_0(\mathbf{r},\mathbf{r}^\prime;\omega)\right]=-\frac{k^3}{15\pi}(x_i-x_i^\prime)\mathbf{I}+\frac{k^3}{60\pi}(\hat{\mathbf{e}}_i \mathbf{R}+\mathbf{R}\hat{\mathbf{e}}_i)+\cdots,\end{equation}
in which $\hat{\mathbf{e}}_i$ represents the unit vector along i-direction. From this, we can recognize that the partial derivatives of $\mathbf{G}_0$ vanishes at the source point by setting $R=0$: 
\begin{equation}\mathrm{Im}\left[\partial_i\mathbf{G}_0(\mathbf{r}_p,\mathbf{r}_p;\omega)\right]=0.\end{equation}
If we further apply $\partial_j^\prime$ on both sides of Eq. \ref{eq:dGi}, one obtains
\begin{equation}\mathrm{Im}\left[\partial_i\partial_j^\prime\mathbf{G}_0(\mathbf{r}_p,\mathbf{r}_p;\omega)\right]=\frac{k^3}{15\pi}\delta_{ij}\mathbf{I}-\frac{k^3}{60\pi}(\hat{\mathbf{e}}_i\hat{\mathbf{e}}_j+\hat{\mathbf{e}}_j\hat{\mathbf{e}}_i),\end{equation}
where $\delta_{ij}$ denotes the Kronecker delta. 
\subsection{Auto-correlation of $\delta \hat{F}_x$}
The fluctuating force along the x-direction $\delta \hat{F}_x$  is described in Eq. 10 of the main text. The auto-correlation of $\delta \hat{F}_x$ can be written as: 
\begin{equation}\begin{split}\langle \delta \hat{F}_x(t)\delta \hat{F}_x(t+\tau) \rangle&=\bar{p}_x(t)\bar{p}_x(t+\tau)\int_0^\infty\int_0^\infty \langle\partial_x\hat{E}_{N_x}(\mathrm{r}_p,\omega)\partial_x\hat{E}_{N_x}^\dagger(\mathrm{r}_p,\omega^\prime)\rangle e^{-i\omega t} e^{i\omega^\prime(t+\tau)} d\omega d\omega^\prime\\&+\bar{p}_x(t) \partial_z\bar{E}_x(t+\tau)\int_0^\infty\int_0^\infty \langle \partial_x\hat{E}_{N_x}(\mathrm{r}_p,\omega)\hat{E}_{N_z}^\dagger(\mathrm{r}_p,\omega^\prime) \rangle \alpha^\ast(\omega^\prime)e^{-i\omega t} e^{i\omega^\prime(t+\tau)} d\omega d\omega^\prime \\&-\bar{p}_x(t)\bar{B}_y(t+\tau)\int_0^\infty\int_0^\infty \langle \partial_x\hat{E}_{N_x}(\mathrm{r}_p,\omega)\hat{E}_{N_z}^\dagger(\mathrm{r}_p,\omega^\prime) \rangle i\omega^\prime \alpha^\ast(\omega^\prime)e^{-i\omega t} e^{i\omega^\prime(t+\tau)} d\omega d\omega^\prime\\&+\partial_z\bar{E}_x(t)\bar{p}_x(t+\tau) \int_0^\infty\int_0^\infty \langle \hat{E}_{N_z}(\mathrm{r}_p,\omega)\partial_x\hat{E}_{N_x}^\dagger(\mathrm{r}_p,\omega^\prime) \rangle \alpha(\omega)e^{-i\omega t} e^{i\omega^\prime(t+\tau)} d\omega d\omega^\prime\\&+\partial_z\bar{E}_x(t)\partial_z\bar{E}_x(t+\tau)\int_0^\infty\int_0^\infty \langle \hat{E}_{N_z}(\mathrm{r}_p,\omega)\hat{E}_{N_z
}^\dagger(\mathrm{r}_p,\omega^\prime) \rangle \alpha(\omega)\alpha^\ast(\omega^\prime)e^{-i\omega t} e^{i\omega^\prime(t+\tau)} d\omega d\omega^\prime\\&-\partial_z\bar{E}_x(t)\bar{B}_y(t+\tau)\int_0^\infty\int_0^\infty \langle \hat{E}_{N_z}(\mathrm{r}_p,\omega)\hat{E}_{N_z
}^\dagger(\mathrm{r}_p,\omega^\prime) \rangle i\omega^\prime \alpha(\omega)\alpha^\ast(\omega^\prime)e^{-i\omega t} e^{i\omega^\prime(t+\tau)} d\omega d\omega^\prime\\&-\bar{B}_y(t)\bar{p}_x(t+\tau) \int_0^\infty\int_0^\infty \langle \hat{E}_{N_z}(\mathrm{r}_p,\omega)\partial_x\hat{E}_{N_x
}^\dagger(\mathrm{r}_p,\omega^\prime) \rangle (-i\omega) \alpha(\omega)e^{-i\omega t} e^{i\omega^\prime(t+\tau)} d\omega d\omega^\prime\\&-\bar{B}_y(t)\partial_z\bar{E}_x(t+\tau) \int_0^\infty\int_0^\infty \langle \hat{E}_{N_z}(\mathrm{r}_p,\omega)\hat{E}_{N_z
}^\dagger(\mathrm{r}_p,\omega^\prime) \rangle (-i\omega) \alpha(\omega)\alpha^\ast(\omega^\prime) e^{-i\omega t} e^{i\omega^\prime(t+\tau)} d\omega d\omega^\prime\\&+\bar{B}_y(t)\bar{B}_y(t+\tau) \int_0^\infty\int_0^\infty \langle \hat{E}_{N_z}(\mathrm{r}_p,\omega)\hat{E}_{N_z
}^\dagger(\mathrm{r}_p,\omega^\prime) \rangle \omega\omega^\prime \alpha(\omega) \alpha^\ast(\omega^\prime) e^{-i\omega t} e^{i\omega^\prime(t+\tau)} d\omega d\omega^\prime.\end{split}\end{equation}
By utilizing the electromagnetic noise correlations given by Eq. 2 of the main text and employing the following relations regarding the Green's function,
\begin{subequations}
\begin{equation}\mathrm{Im}\left[\partial_x\partial_x^\prime G_{0xx}(\mathbf{r}_p,\mathbf{r}_p;\omega)\right]=\frac{\omega^3}{30\pi c^3},\end{equation}
\begin{equation}\mathrm{Im}\left[ G_{0zz}(\mathbf{r}_p,\mathbf{r}_p;\omega)\right]=\frac{\omega}{6\pi c},\end{equation}
\begin{equation}\mathrm{Im}\left[\partial_x G_{0xz}(\mathbf{r}_p,\mathbf{r}_p;\omega)\right]=\mathrm{Im}\left[\partial_x^\prime G_{0zx}(\mathbf{r}_p,\mathbf{r}_p;\omega)\right]=0,\end{equation}
\end{subequations}
we can simplify the auto-correlation of $\delta \hat{F}_x$ into:
\begin{equation}\begin{split}\langle \delta \hat{F}_x(t)\delta \hat{F}_x(t+\tau) \rangle&=\bar{p}_x(t)\bar{p}_x(t+\tau)\int_0^\infty \frac{\hbar\mu_0}{\pi}\times\frac{\omega^5}{30\pi c^3} e^{i\omega \tau}d\omega\\&+\partial_z \bar{E}_x(t)\partial_z \bar{E}_x(t+\tau)\int_0^\infty \frac{\hbar\mu_0}{\pi}\times\frac{\omega^3}{6\pi c}|\alpha(\omega)|^2 e^{i\omega \tau}d\omega \\&-\partial_z \bar{E}_x(t) \bar{B}_y(t+\tau)\int_0^\infty \frac{\hbar\mu_0}{\pi}\times\frac{i\omega^4}{6\pi c}|\alpha(\omega)|^2 e^{i\omega \tau}d\omega \\&-\bar{B}_y(t)\partial_z \bar{E}_x(t+\tau)\int_0^\infty \frac{\hbar\mu_0}{\pi}\times\frac{i\omega^4}{6\pi c}|\alpha(\omega)|^2 e^{i\omega \tau}d\omega\\ &+\bar{B}_y(t) \bar{B}_y(t+\tau)\int_0^\infty \frac{\hbar\mu_0}{\pi}\times\frac{\omega^5}{6\pi c}|\alpha(\omega)|^2 e^{i\omega \tau}d\omega.\end{split}\end{equation}
By substituting the following relations, which are obtained after neglecting high oscillating terms:
\begin{subequations}
\begin{equation}\bar{p}_x(t)\bar{p}_x(t+\tau)=\frac{1}{2}\mathrm{Re}[E_0^2|\alpha(\omega_0)|^2e^{i\omega_0 \tau}],\end{equation}
\begin{equation}\partial_z \bar{E}_x(t)\partial_z \bar{E}_x(t+\tau)=\frac{1}{2}\mathrm{Re}[\frac{\omega_0^2}{c^2}E_0^2e^{i\omega_0 \tau}],\end{equation}
\begin{equation}\partial_z \bar{E}_x(t) \bar{B}_y(t+\tau)=\frac{1}{2}\mathrm{Re}[\frac{i\omega_0}{c^2}E_0^2e^{i\omega_0 \tau}],\end{equation}
\begin{equation}\bar{B}_y(t)\partial_z \bar{E}_x(t+\tau) =\frac{1}{2}\mathrm{Re}[\frac{-i\omega_0}{c^2}E_0^2e^{i\omega_0 \tau}],\end{equation}
\begin{equation}\bar{B}_y(t) \bar{B}_y(t+\tau)=\frac{1}{2}\mathrm{Re}[\frac{E_0^2}{c^2}e^{i\omega_0 \tau}],\end{equation}
\end{subequations}
we obtain
\begin{equation}\begin{split}\langle \delta \hat{F}_x(t)\delta \hat{F}_x(t+\tau) \rangle&=\frac{\hbar\mu_0}{2\pi}\mathrm{Re}\left[|\alpha(\omega_0)|^2E_0^2 e^{i\omega_0\tau}\right]\int_0^\infty \frac{\omega^5}{30\pi c^3} e^{i\omega \tau}d\omega\\&+\frac{\hbar\mu_0}{2\pi}\mathrm{Re}\left[\frac{\omega_0^2}{c^2}E_0^2 e^{i\omega_0\tau}\right]\int_0^\infty \frac{\omega^3}{6\pi c}|\alpha(\omega)|^2 e^{i\omega \tau}d\omega \\&-\frac{\hbar\mu_0}{\pi}\mathrm{Re}\left[\frac{i\omega_0}{c^2}E_0^2 e^{i\omega_0\tau}\right]\int_0^\infty \frac{i\omega^4}{6\pi c}|\alpha(\omega)|^2 e^{i\omega \tau}d\omega \\ &+\frac{\hbar\mu_0}{2\pi}\mathrm{Re}\left[\frac{E_0^2}{c^2}e^{i\omega_0\tau}\right]\int_0^\infty \frac{\omega^5}{6\pi c}|\alpha(\omega)|^2 e^{i\omega \tau}d\omega.\end{split}\end{equation}

\subsection{Auto-correlation of $\delta \hat{F}_y$}
The fluctuating force along the y-direction $\delta \hat{F}_y$  is described in Eq. 10 of the main text. The auto-correlation of $\delta \hat{F}_y$ can be written as
\begin{equation}\begin{split}&\langle \delta \hat{F}_y(t)\delta \hat{F}_y(t+\tau) \rangle=\\ &\bar{p}_x(t)\bar{p}_x(t+\tau)\int_0^\infty\int_0^\infty \langle\partial_x\hat{E}_{N_y}(\mathrm{r}_p,\omega)\partial_x\hat{E}_{N_y}^\dagger(\mathrm{r}_p,\omega^\prime)\rangle e^{-i\omega t} e^{i\omega^\prime(t+\tau)} d\omega d\omega^\prime-\\& \bar{p}_x(t)\partial_t\bar{p}_x(t+\tau)\int_0^\infty\int_0^\infty \langle\partial_x\hat{E}_{N_y}(\mathrm{r}_p,\omega)\left[\partial_x\hat{E}_{N_y}^\dagger(\mathrm{r}_p,\omega^\prime)-\partial_y\hat{E}_{N_x}^\dagger(\mathrm{r}_p,\omega^\prime)\right]\rangle\frac{1}{-i\omega^\prime} e^{-i\omega t} e^{i\omega^\prime(t+\tau)} d\omega d\omega^\prime- \\ &\partial_t\bar{p}_x(t)\bar{p}_x(t+\tau)\int_0^\infty\int_0^\infty \langle\left[\partial_x\hat{E}_{N_y}(\mathrm{r}_p,\omega^\prime)-\partial_y\hat{E}_{N_x}(\mathrm{r}_p,\omega^\prime)\right]\partial_x\hat{E}_{N_y}^\dagger(\mathrm{r}_p,\omega)\rangle \frac{1}{i\omega} e^{-i\omega t} e^{i\omega^\prime(t+\tau)} d\omega d\omega^\prime+\\ &\partial_t\bar{p}_x(t) \partial_t\bar{p}_x(t+\tau)\int_0^\infty\int_0^\infty \langle\left[\partial_x\hat{E}_{N_y}(\mathrm{r}_p,\omega^\prime)-\partial_y\hat{E}_{N_x}(\mathrm{r}_p,\omega^\prime)\right]\left[\partial_x\hat{E}_{N_y}^\dagger(\mathrm{r}_p,\omega)-\partial_y\hat{E}_{N_x}^\dagger(\mathrm{r}_p,\omega)\right]\rangle \frac{1}{\omega \omega^\prime} e^{-i\omega t} e^{i\omega^\prime(t+\tau)} d\omega d\omega^\prime. \end{split}\end{equation}
By utilizing the electromagnetic noise correlations given by Eq. 2 of the main text and employing the following relations regarding the Green's function
\begin{equation}\mathrm{Im}\left[\partial_x\partial_x^\prime G_{0yy}(\mathbf{r}_p,\mathbf{r}_p;\omega)\right]=\mathrm{Im}\left[ \partial_y \partial_y^\prime G_{0xx}(\mathbf{r}_p,\mathbf{r}_p;\omega)\right]=\frac{\omega^3}{15\pi c^3},\end{equation}
\begin{equation}\mathrm{Im}\left[\partial_x \partial_y^\prime G_{0yx}(\mathbf{r}_p,\mathbf{r}_p;\omega)\right]=\mathrm{Im}\left[\partial_y \partial_x^\prime G_{0xy}(\mathbf{r}_p,\mathbf{r}_p;\omega)\right]=\frac{-\omega^3}{60\pi c^3},\end{equation}
we can simplify the auto-correlation of $\delta \hat{F}_y$ into:
\begin{equation}\begin{split}\langle \delta \hat{F}_y(t)\delta \hat{F}_y(t+\tau) \rangle= &\bar{p}_x(t)\bar{p}_x(t+\tau)\int_0^\infty \frac{\hbar\mu_0}{2\pi}\times\frac{\omega^5}{15\pi c^3} e^{i\omega \tau}d\omega\\&-\bar{p}_x(t)\partial_t\bar{p}_x(t+\tau)\int_0^\infty \frac{\hbar\mu_0}{\pi}\times\frac{i\omega^4}{12\pi c^3} e^{i\omega \tau}d\omega \\&+\partial_t\bar{p}_x(t)\bar{p}_x(t+\tau)\int_0^\infty \frac{\hbar\mu_0}{\pi}\times\frac{i\omega^4}{12\pi c^3} e^{i\omega \tau}d\omega \\ &+\partial_t\bar{p}_x(t)\partial_t\bar{p}_x(t+\tau)\int_0^\infty \frac{\hbar\mu_0}{2\pi}\times\frac{\omega^3}{6\pi c^3} e^{i\omega \tau}d\omega.\end{split}\end{equation}
By substituting the following relations, which are obtained after neglecting high oscillating terms:
\begin{equation}\bar{p}_x(t)\bar{p}_x(t+\tau)=\frac{1}{2}\mathrm{Re}[E_0^2|\alpha(\omega_0)|^2e^{i\omega_0 \tau}],\end{equation}
\begin{equation}\bar{p}_x(t)\partial_t\bar{p}_x(t+\tau)=\frac{1}{2}\mathrm{Re}[i\omega_0 E_0^2|\alpha(\omega_0)|^2e^{i\omega_0 \tau}],\end{equation}
\begin{equation}\partial_t\bar{p}_x(t)\bar{p}_x(t+\tau)=\frac{1}{2}\mathrm{Re}[-i\omega_0E_0^2|\alpha(\omega_0)|^2e^{i\omega_0 \tau}],\end{equation}
\begin{equation}\partial_t\bar{p}_x(t)\partial_t\bar{p}_x(t+\tau)=\frac{1}{2}\mathrm{Re}[\omega_0^2 E_0^2|\alpha(\omega_0)|^2e^{i\omega_0 \tau}],\end{equation}
one obtains
\begin{equation}\begin{split}\langle \delta \hat{F}_y(t)\delta \hat{F}_y(t+\tau) \rangle= &\frac{\hbar\mu_0}{2\pi}\mathrm{Re}\left[|\alpha(\omega_0)|^2E_0^2 e^{i\omega_0\tau}\right]\int_0^\infty \frac{\omega^5}{15\pi c^3} e^{i\omega \tau}d\omega\\&-\frac{\hbar\mu_0}{\pi}\mathrm{Re}\left[i\omega_0 |\alpha(\omega_0)|^2E_0^2 e^{i\omega_0\tau}\right]\int_0^\infty \frac{i\omega^4}{12\pi c^3} e^{i\omega \tau}d\omega \\ &+\frac{\hbar\mu_0}{2\pi}\mathrm{Re}\left[\omega_0^2|\alpha(\omega_0)|^2E_0^2e^{i\omega_0\tau}\right]\int_0^\infty \frac{\omega^3}{6\pi c^3} e^{i\omega \tau}d\omega.\end{split}\end{equation}

\subsection{Auto-correlation of $\delta \hat{F}_z$}
The fluctuating force along the z-direction $\delta \hat{F}_z$  is described in Eq. 10 of the main text. The auto-correlation of $\delta \hat{F}_z$ can be written as
\begin{equation}\begin{split}&\langle \delta \hat{F}_z(t)\delta \hat{F}_z(t+\tau) \rangle=\\ &\bar{p}_x(t)\bar{p}_x(t+\tau)\int_0^\infty\int_0^\infty \langle\partial_x\hat{E}_{N_z}(\mathrm{r}_p,\omega)\partial_x\hat{E}_{N_z}^\dagger(\mathrm{r}_p,\omega^\prime)\rangle e^{-i\omega t} e^{i\omega^\prime(t+\tau)} d\omega d\omega^\prime+\\& \bar{p}_x(t)\partial_t\bar{p}_x(t+\tau)\int_0^\infty\int_0^\infty \langle\partial_x\hat{E}_{N_z}(\mathrm{r}_p,\omega)\left[\partial_z\hat{E}_{N_x}^\dagger(\mathrm{r}_p,\omega^\prime)-\partial_x\hat{E}_{N_z}^\dagger(\mathrm{r}_p,\omega^\prime)\right]\rangle \frac{1}{-i\omega^\prime} e^{-i\omega t} e^{i\omega^\prime(t+\tau)} d\omega d\omega^\prime+\\&\bar{p}_x(t)\bar{B}_y(t+\tau)\int_0^\infty\int_0^\infty \langle\partial_x\hat{E}_{N_z}(\mathrm{r}_p,\omega)\hat{E}_{N_x}^\dagger(\mathrm{r}_p,\omega^\prime)\rangle i\omega^\prime \alpha^\ast(\omega^\prime) e^{-i\omega t} e^{i\omega^\prime(t+\tau)} d\omega d\omega^\prime+\\& \partial_t\bar{p}_x(t)\bar{p}_x(t+\tau)\int_0^\infty\int_0^\infty \langle \left[\partial_z\hat{E}_{N_x}(\mathrm{r}_p,\omega)-\partial_x\hat{E}_{N_z}(\mathrm{r}_p,\omega)\right]\partial_x\hat{E}_{N_z}^\dagger(\mathrm{r}_p,\omega^\prime)\rangle \frac{1}{i\omega} e^{-i\omega t} e^{i\omega^\prime(t+\tau)} d\omega d\omega^\prime+\\& \partial_t\bar{p}_x(t) \partial_t\bar{p}_x(t+\tau)\int_0^\infty\int_0^\infty \langle\left[\partial_z\hat{E}_{N_x}(\mathrm{r}_p,\omega)-\partial_x\hat{E}_{N_z}(\mathrm{r}_p,\omega)\right]\left[\partial_z\hat{E}_{N_x}^\dagger(\mathrm{r}_p,\omega^\prime)-\partial_x\hat{E}_{N_z}^\dagger(\mathrm{r}_p,\omega^\prime)\right]\rangle \frac{1}{\omega \omega^\prime} e^{-i\omega t} e^{i\omega^\prime(t+\tau)} d\omega d\omega^\prime+\\& \partial_t\bar{p}_x(t)\bar{B}_y(t+\tau)\int_0^\infty\int_0^\infty \langle\left[\partial_z\hat{E}_{N_x}(\mathrm{r}_p,\omega)-\partial_x\hat{E}_{N_z}(\mathrm{r}_p,\omega)\right]\hat{E}_{N_x}^\dagger(\mathrm{r}_p,\omega^\prime)\rangle \frac{\omega^\prime}{\omega} \alpha^\ast(\omega^\prime) e^{-i\omega t} e^{i\omega^\prime(t+\tau)} d\omega d\omega^\prime+\\ &\bar{B}_y(t)\bar{p}_x(t+\tau)\int_0^\infty\int_0^\infty \langle\hat{E}_{N_x}(\mathrm{r}_p,\omega)\partial_x\hat{E}_{N_z}^\dagger(\mathrm{r}_p,\omega^\prime)\rangle (-i\omega)\alpha(\omega) e^{-i\omega t} e^{i\omega^\prime(t+\tau)} d\omega d\omega^\prime+\\ &\bar{B}_y(t)\partial_t\bar{p}_x(t+\tau)\int_0^\infty\int_0^\infty \langle\hat{E}_{N_x}(\mathrm{r}_p,\omega)\left[\partial_z\hat{E}_{N_x}^\dagger(\mathrm{r}_p,\omega^\prime)-\partial_x\hat{E}_{N_z}^\dagger(\mathrm{r}_p,\omega^\prime)\right]\rangle \frac{\omega}{\omega^\prime} \alpha(\omega) e^{-i\omega t} e^{i\omega^\prime(t+\tau)} d\omega d\omega^\prime+\\ &\bar{B}_y(t)\bar{B}_y(t+\tau)\int_0^\infty\int_0^\infty \langle\hat{E}_{N_x}(\mathrm{r}_p,\omega)\hat{E}_{N_x}^\dagger(\mathrm{r}_p,\omega^\prime)\rangle \omega \omega^\prime \alpha(\omega)\alpha^\ast(\omega^\prime) e^{-i\omega t} e^{i\omega^\prime(t+\tau)} d\omega d\omega^\prime. \end{split}\end{equation}
By utilizing the electromagnetic noise correlations given by Eq. 2 of the main text and employing the following relations regarding the Green's function
\begin{subequations}
\begin{equation}\mathrm{Im}\left[\partial_x\partial_x^\prime G_{0zz}(\mathbf{r}_p,\mathbf{r}_p;\omega)\right]=\mathrm{Im}\left[ \partial_z \partial_z^\prime G_{0xx}(\mathbf{r}_p,\mathbf{r}_p;\omega)\right]=\frac{\omega^3}{15\pi c^3},\end{equation}
\begin{equation}\mathrm{Im}\left[\partial_x \partial_z^\prime G_{0zx}(\mathbf{r}_p,\mathbf{r}_p;\omega)\right]=\mathrm{Im}\left[\partial_z \partial_x^\prime G_{0xz}(\mathbf{r}_p,\mathbf{r}_p;\omega)\right]=\frac{-\omega^3}{60\pi c^3},\end{equation}
\begin{equation}\mathrm{Im}\left[ G_{0xx}(\mathbf{r}_p,\mathbf{r}_p;\omega)\right]=\frac{\omega}{6\pi c},\end{equation}
\begin{equation}\mathrm{Im}\left[ \partial_x G_{0zx}(\mathbf{r}_p,\mathbf{r}_p;\omega)\right]=\mathrm{Im}\left[ \partial_z G_{0xx}(\mathbf{r}_p,\mathbf{r}_p;\omega)\right]=\mathrm{Im}\left[ \partial_x^\prime G_{0xz}(\mathbf{r}_p,\mathbf{r}_p;\omega)\right]=0,\end{equation}
\end{subequations}
we can simplify the auto-correlation of $\delta \hat{F}_z$ into:
\begin{equation}\begin{split}\langle \delta \hat{F}_z(t)\delta \hat{F}_z(t+\tau) \rangle= &\bar{p}_x(t)\bar{p}_x(t+\tau)\int_0^\infty \frac{\hbar\mu_0}{2\pi}\times\frac{\omega^5}{15\pi c^3} e^{i\omega \tau}d\omega\\-&\bar{p}_x(t)\partial_t\bar{p}_x(t+\tau)\int_0^\infty \frac{\hbar\mu_0}{\pi}\times\frac{i\omega^4}{12\pi c^3} e^{i\omega \tau}d\omega\\+&\partial_t\bar{p}_x(t)\bar{p}_x(t+\tau)\int_0^\infty \frac{\hbar\mu_0}{\pi}\times\frac{i\omega^4}{12\pi c^3} e^{i\omega \tau}d\omega\\+ &\partial_t\bar{p}_x(t)\partial_t\bar{p}_x(t+\tau)\int_0^\infty \frac{\hbar\mu_0}{\pi}\times\frac{\omega^3}{6\pi c^3} e^{i\omega \tau}d\omega\\+&\bar{B}_y(t) \bar{B}_y(t+\tau)\int_0^\infty \frac{\hbar\mu_0}{\pi}|\alpha(\omega_0)|^2\frac{\omega^5}{6\pi c} e^{i\omega \tau}d\omega. \end{split}\end{equation}
By substituting the following relations, which are obtained after neglecting high oscillating terms:
\begin{subequations}
\begin{equation}\bar{p}_x(t)\bar{p}_x(t+\tau)=\frac{1}{2}\mathrm{Re}[E_0^2|\alpha(\omega_0)|^2e^{i\omega_0 \tau}],\end{equation}
\begin{equation}\bar{p}_x(t)\partial_t\bar{p}_x(t+\tau)=\frac{1}{2}\mathrm{Re}[i\omega_0 E_0^2|\alpha(\omega_0)|^2e^{i\omega_0 \tau}],\end{equation}
\begin{equation}\partial_t\bar{p}_x(t)\bar{p}_x(t+\tau)=\frac{1}{2}\mathrm{Re}[-i\omega_0 E_0^2|\alpha(\omega_0)|^2e^{i\omega_0 \tau}],\end{equation}
\begin{equation}\partial_t\bar{p}_x(t)\partial_t\bar{p}_x(t+\tau)=\frac{1}{2}\mathrm{Re}[\omega_0^2 E_0^2|\alpha(\omega_0)|^2e^{i\omega_0 \tau}],\end{equation}
\begin{equation}\bar{B}_y(t) \bar{B}_y(t+\tau)=\frac{1}{2}\mathrm{Re}[\frac{E_0^2}{c^2}e^{i\omega_0 \tau}],\end{equation}
\end{subequations}
one obtains
\begin{equation}\begin{split}\langle \delta \hat{F}_z(t)\delta \hat{F}_z(t+\tau) \rangle= &\frac{\hbar\mu_0}{2\pi}\mathrm{Re}\left[|\alpha(\omega_0)|^2E_0^2 e^{i\omega_0\tau}\right]\int_0^\infty \frac{\omega^5}{15\pi c^3} e^{i\omega \tau}d\omega\\-&\frac{\hbar\mu_0}{\pi}\mathrm{Re}\left[i\omega_0 |\alpha(\omega_0)|^2E_0^2 e^{i\omega_0\tau}\right]\int_0^\infty \frac{i\omega^4}{12\pi c^3} e^{i\omega \tau}d\omega\\+ &\frac{\hbar\mu_0}{2\pi}\mathrm{Re}\left[\omega_0^2|\alpha(\omega_0)|^2E_0^2e^{i\omega_0\tau}\right]\int_0^\infty \frac{\omega^3}{6\pi c^3} e^{i\omega \tau}d\omega\\+&\frac{\hbar\mu_0}{2\pi}\mathrm{Re}\left[\frac{E_0^2}{c^2}e^{i\omega_0\tau}\right]\int_0^\infty |\alpha(\omega_0)|^2\frac{\omega^5}{6\pi c} e^{i\omega \tau}d\omega. \end{split}\end{equation}

